\documentclass[12pt]{article}

\begin{document}

\title{Quantum-mechanical probability from the symmetries of
       two-state systems\\[10mm]}
\vspace{10mm}
\author{{\bf L. Polley}\\[5mm]
Physics Department\\Oldenburg University\\26111 Oldenburg\\Germany\\[10mm]}
\maketitle
\begin{abstract}
In 1989, Deutsch gave a basic physical explanation of 
why quantum-mechanical probabilities are squares of amplitudes. 
Essentially, a general state vector is transformed into a 
highly symmetric equal-amplitude superposition. 
The argument was recently elaborated and publicised by DeWitt.
It has remained incomplete, however, inasmuch as both authors anticipate 
the usual normalization (sum of amplitudes squared) of state vectors. 
In the present paper, a thought experiment is devised in which Deutsch's 
idea is demonstrated independently of the normalization, exploiting 
further symmetries instead.
\end{abstract}
\thispagestyle{empty}

\newpage\noindent
According to the standard Born statistical interpretation of a state vector 
$$
   |\psi\rangle = \sum_i \psi_i |i\rangle \qquad \qquad
                                                        \sum_k |\psi_k|^2 = 1
$$ 
the $i$th eigenvalue of an observable is measured with probability 
$|\psi_i|^2$. While there is no generally accepted answer as
to the origin of the stochasticity, the values of the probabilities can  
be deduced from a variety of assumptions. The simplest way is to define a
quantum state as a linear expectation-value functional over the algebra of 
observables \cite{Emch,Peres}. Thus one starts out from 
\begin{equation}   \label{A+B}
   \langle A + B \rangle = \langle A \rangle + \langle B \rangle  
   \hspace*{20mm} \mbox{(not used in this paper)}
\end{equation}
Such an equation can be taken to define  ``correspondence'' \cite{Ulfbeck} 
but its physical interpretation is not unproblematic \cite{Peres}.
$A$ and $B$ are formal representations of apparatuses, and it is 
hard to tell what kind of apparatus is represented by $A+B$ if the
summands are as different as a particle's momentum and position.  

A celebrated mathematical result on quantum-mechanical probabilities 
is due to Gleason \cite{Gleason1}.
If normalized probabilities are to be assigned to the eigenvectors 
of each hermitian operator, which is what a quantum state is expected to do,
then the only possibility is the amplitudes-squared prescription. 
Unfortunately, the existing proofs of Gleason's theorem 
(cf.\ \cite{Pitowsky,Peres}) are not easily received by many physicists.

In Deutsch's approach \cite{Deutsch1,Deutsch2,DeWitt} only 
basic mathematics is involved. Moreover, sums of operators as on the
{\sc lhs} of (\ref{A+B}) can be avoided; only superpositions of state 
vectors are required. DeWitt \cite{DeWitt} conceals this latter advantage 
by presenting the argument in terms of sums of projection operators, 
while Deutsch \cite{Deutsch2} uses state vectors only. 
However, both authors anticipate in a crucial way the amplitudes-squared
normalization of physical state vectors. The problem with this is that the
standard normalization is physically motivated by the amplitudes-squared form 
of the probabilities.  

In the present paper, normalizations are implicit in the unitarity of time 
evolutions. The latter can be inferred from something weaker than unitarity: 
from the assumption that none of the state vectors ``decay'' to the null vector. 
This would also be consistent with probabilities equal to the absolute values 
(unsquared) of amplitudes, hence it is a weaker assumption. In conjunction with 
symmetries of certain two-state subsystems, however, full unitarity is 
recovered automatically.

Following Deutsch \cite{Deutsch2}, let us consider a superposition state 
of the special form
\begin{equation}     \label{DeutschSup}
    |\psi\rangle = \sqrt{\frac{m}{m+n}}~|A\rangle + \sqrt{\frac{n}{m+n}}~|B\rangle 
\end{equation}
Let this be coupled to an auxilliary $m+n+1$-state system, and let $|A\rangle$ and
$|B\rangle$ be substituted by {\em normalized} superpositions according to
\begin{eqnarray}   \label{DeutschSubst1} 
  |A\rangle \rightarrow |A\rangle |0\rangle & \stackrel{S}{\longrightarrow} & 
                            \sqrt{\frac1m}\sum_{i=1}^m |A\rangle |i\rangle \\
  |B\rangle \rightarrow |B\rangle |0\rangle & \stackrel{S}{\longrightarrow} & 
                            \sqrt{\frac1{n}}\sum_{i=m+1}^{m+n} |B\rangle |i\rangle 
\label{DeutschSubst2}
\end{eqnarray}
Deutsch \cite{Deutsch2} motivates $S$ by decision-theoretic 
substitutability; DeWitt \cite{DeWitt} devises an observable whose 
expectation value involves $S$.  
The substitution preserves the properties $A$ and $B$, but when inserted in 
(\ref{DeutschSup}) it results in an equal-amplitude superposition of the form
\begin{equation}     \label{equal}
    \frac1{\sqrt{m+n}} \sum_{M=1}^{m+n} |M\rangle    
\end{equation}
Due to the permutational symmetry (see also discussion of (\ref{sumAi}) below) 
the probability for detecting an $|M\rangle$ state is $(m+n)^{-1}$. Thus the 
probability for property $A$ is $(m+n)^{-1}m$, and for $B$ it is $(m+n)^{-1}n$.

In order to avoid anticipating the normalization,
$S$ is interpreted here as the time evolution of an apparatus 
capable of spatially separating internal states of an atom.
The factors of $1/\sqrt{m}$ and $1/\sqrt{n}$ then arise automatically.
It is reassuring to note that a variety of state-separating apparatuses 
can be realized experimentally \cite{SternG,Monroe}.

Consider an arrangement of $m+n+1$ cavities, all of the same shape, connected
by channels as indicated for $m=3$ and $n=2$ in Figure 1. 
\begin{figure}
\unitlength 1.0cm
\hspace*{15mm}
\begin{picture}(10,6)
\thicklines
\put(5.0,3.0){\circle{2}}   \put(4.5,2.9){$|\psi\rangle|0\rangle$}
\put(1.0,3.0){\circle{2}}   \put(0.5,2.90){$|A\rangle|2\rangle$}
\put(1.67,5.22){\circle{2}} \put(1.15,5.12){$|A\rangle|1\rangle$}
\put(1.67,0.78){\circle{2}} \put(1.15,0.68){$|A\rangle|3\rangle$}
\put(4.30,3.00){\line(-1,+0){2.60}}
\put(4.42,3.39){\line(-3,+2){2.16}}
\put(4.42,2.61){\line(-3,-2){2.16}}
\put(5.68,3.17){\line(4,+1){2.52}}
\put(8.88,3.97){\circle{2}} \put(8.35,3.87){$|B\rangle|5\rangle$}
\put(5.68,2.83){\line(4,-1){2.52}}
\put(8.88,2.03){\circle{2}} \put(8.35,1.93){$|B\rangle|4\rangle$}
\end{picture}
\caption{\protect\small A device for deducing the probabilities $3/5$ 
and $2/5$ for $A$ and $B$, respectively, from the state vector 
$|\psi\protect\rangle =
 \protect\sqrt{3}~|A\protect\rangle + \protect\sqrt{2}~|B\protect\rangle$.
The normalization of $|\psi\protect\rangle$ is not essential for the argument.}
\end{figure}
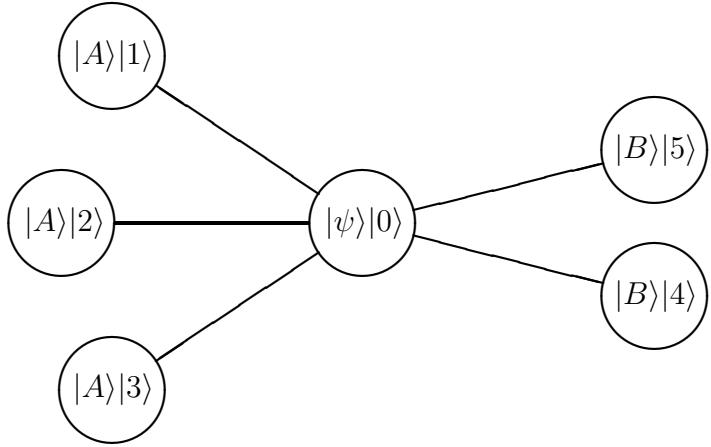
Let the states $|A\rangle$ 
and $|B\rangle$ correspond to internal states of an atom. If the atom is placed 
in cavity $i$, its total state is $|A\rangle |i\rangle$ or $|B\rangle |i\rangle$, 
etc. Let us assume that channels $1,\ldots,m$ can be passed by the atom in state 
$|A\rangle$ only, channels $m+1,\ldots,m+n$ in state $|B\rangle$ only, and that 
channels can be closed individually. 

Why cavities? Their point is to enable an individual treatment of parts of a
wavefunction. If summand $|i\rangle$ resides in a disconnected cavity it 
is screened from other summands $|j\rangle\neq|i\rangle$. We shall use this  
(a)  to put into storage a summand that already has a desired form, and  
(b) to change the complex phase or internal state of a summand.     

A particularly symmetric situation arises if we close all channels but one,
connecting $|0\rangle$ to some $|i\rangle$. The permutation of $|0\rangle$ and 
$|i\rangle$ is then a symmetry of the time evolution. An atom in the initial state 
$|A\rangle |0\rangle$ will evolve, after a time interval $\tau$,
into a superposition 
\begin{equation}      \label{0hop}  
    \alpha |A\rangle |0\rangle + \beta |A\rangle |i\rangle 
\end{equation} 
where $\alpha$ and $\beta$ are complex numbers. 
By exchanging the roles of $|0\rangle$ and $|i\rangle$ we obtain from an 
initial state $|A\rangle |i\rangle$
\begin{equation}      \label{ihop}  
   \beta |A\rangle |0\rangle + \alpha |A\rangle |i\rangle 
\end{equation} 
There are two stationary cavity states, $|\pm\rangle = |0\rangle\pm|i\rangle$, 
for which the time evolution takes the form (omitting the $|A\rangle$ factor 
for the moment)
\begin{equation}      \label{station}
    |\pm\rangle \longrightarrow (\alpha\pm\beta) ~ |\pm\rangle
\end{equation} 
We now postulate the system to be {\em stable} in the sense that none of the 
state vectors tend to zero for $t\to\pm\infty$. Repeated application of 
(\ref{station}) then implies both
$|\alpha+\beta|^2=1$ and $|\alpha-\beta|^2=1$, which in turn implies
$$
    |\alpha|^2 + |\beta|^2 = 1     \qquad 
    \alpha \beta^* + \alpha^* \beta = 0
$$
Thus the time evolution matrix 
$$
   U(\tau) = \left( \begin{array}{cc} \alpha(\tau) & \beta(\tau)  \\ 
                                      \beta(\tau)  & \alpha(\tau) 
                     \end{array} \right) 
$$
is unitary automatically for any $\tau$. Its explicit dependence on
$\tau$ can be seen from the group property $U(\tau+\tau')=U(\tau)U(\tau')$
(by considering infinitesimal $\tau'$ and integrating up, for example):
\begin{equation}    \label{Utau}
   U(\tau) = e^{i\epsilon\tau}
     \left( \begin{array}{cc} \cos\omega\tau & i\sin\omega\tau \\ 
             i\sin\omega\tau & \cos\omega\tau \end{array} \right)
\end{equation}
Parameters $\epsilon$ and $\omega$ are real numbers defined by 
$e^{i\epsilon\tau}\cos\omega\tau=\alpha$. Thus we recover the well-known
time evolution of a symmetric two-state system {\em without} anticipating  
conservation of probability.  

We now use the time evolution through individual channels in order to 
produce an equal-amplitude superposition from (\ref{DeutschSup}). 
The iterative step is
\begin{equation}  \label{iterative}
   \sqrt{k} |A\rangle |0\rangle \longrightarrow  \sqrt{k-1}|A\rangle |0\rangle 
                                           + |A\rangle |i\rangle
\end{equation}
which is accomplished by opening channel $i$ exclusively, for a time interval 
$\tau_k$ determined by
$$
    \cot \omega\tau_k = \sqrt{k-1}
$$
Thus we produce a unit-amplitude contribution in the $i$th cavity.  
In fact, we are allowed to consider simplified, non-normalized state vectors 
here because it will be the {\em equality}\/ of amplitudes in the various 
cavities that matters. 

Starting out from the simplified state vector 
$$
    \sqrt{m} |A\rangle |0\rangle + \sqrt{n} |B\rangle |0\rangle  
$$
we open up channels $1,\ldots,m$ for a time interval $\tau_{m}$, 
$\tau_{m-1}$, \ldots, $\tau_{1}$, respectively. This brings down the
amplitude of the central state $|A\rangle |0\rangle$ from $\sqrt{m}$ to
$\sqrt{m-1}$, $\sqrt{m-2}$, \ldots, $0$ while the amplitude of the central state 
$|B\rangle |0\rangle$ in the superposition is not affected. 
In exchange for $\sqrt{m}|A\rangle |0\rangle$ we successively obtain terms 
$|A\rangle |1\rangle$, $|A\rangle |2\rangle$, \ldots, $|A\rangle |m\rangle$
in the superposition, all with a unit amplitude. 
Analogously, to deal with internal state $|B\rangle$ we open up channels 
$m+1$ to $m+n$ for time intervals 
$\tau_{n}$, $\tau_{n-1}$, \ldots, $\tau_{1}$, respectively.
In exchange for $\sqrt{n}|B\rangle |0\rangle$ we obtain terms 
$|B\rangle |m+1\rangle$, $|B\rangle |m+2\rangle$, \ldots, $|B\rangle |m+n\rangle$
with a unit amplitude in the superposition. 
It should be stressed again that in all these steps we do not change properties 
$A$ and $B$, we only separate them spatially. 

According to the discussion so far, we have arrived at a state vector
\begin{equation}    \label{sumM}
     \sum_{i=1}^m |A\rangle |i\rangle + \sum_{i=m+1}^{m+n} |B\rangle |i\rangle 
\end{equation}
However, we have neglected the phase factors that arise from the 
$e^{i\epsilon\tau}$ of equation (\ref{Utau}). Moreover, there could be
additional phase factors from the evolution within a cavity during times 
of disconnection. Hence we should rather discuss the more general expression 
$$
     \sum_{i=1}^m e^{i\varphi_i} |A\rangle |i\rangle 
   + \sum_{i=m+1}^{m+n} e^{i\varphi_i} |B\rangle |i\rangle 
$$
But phase factors for individual cavities do not pose a problem phys\-ics\-wise.
The relation between energies and rotating phase factors, a {\em non-statistical} 
axiom of quantum mechanics, implies that we can get rid of the $e^{i\varphi_i}$ by 
temporarily increasing the potential energy of the atom (gravitationally, e.g.) in 
a particular cavity. Hence it actually suffices to consider expression (\ref{sumM}).  

For the assignment of probability $1/n$ to each of the cavity states of  
(\ref{sumM}) we must bring out the permutational 
symmetry among all $m+n$ states more clearly. Again it is helpful to have 
the internal atomic states $A$ and $B$ separated in space.
From the ``experimental'' procedure we know that in cavities $m+1,\ldots,m+n$
the internal state is necessarily $|B\rangle$. To these cavities we now apply 
a $\pi$ pulse so as to rotate $|B\rangle$ into $|A\rangle$. 
Thus we finally arrive at a state of the form    
\begin{equation}    \label{sumAi}
        \sum_{i=1}^{m+n} |A\rangle \, |i\rangle  
\end{equation}
whose permutational symmetry with respect to {\em all} cavities is obvious.
For example, we could swap the contents of cavities $i$ and $j$ without changing
the state vector. Thus the probability for detecting the atom in a particular 
cavity is $1/n$.
From the conduct of the experiment it follows what this means for the
probabilities of $A$ and $B$.

The argument is easily extended to superpositions involving more than two 
internal states of an atom,
$$
   |\psi\rangle = \sum_i \sqrt{n_i} \, |A_i\rangle
$$ 
Successively, each component $\sqrt{n_i}\,|A_i\rangle$ is transformed by
$n_i$ applications of (\ref{iterative}) into $n_i$ unit-amplitude terms 
of a grand superposition analogous to (\ref{sumM}).

\end{document}